\documentclass[a4paper,twoside]{article}

\usepackage{epsfig,subcaption,calc}
\usepackage{amssymb,amstext,amsmath,amsthm}
\usepackage{multicol,multirow,dblfloatfix}
\usepackage{pslatex,apalike}
\usepackage[utf8]{inputenc}
\usepackage{graphicx,xspace,booktabs}
\usepackage{algorithm,algpseudocode}
\usepackage{xcolor,soul}
\usepackage{caption,subcaption}
\usepackage[inline]{enumitem}
\usepackage{hyperref}
\hypersetup{
  colorlinks=true, linkcolor=black, citecolor=black, urlcolor=black,
  pdftitle={Property Inference Attacks on Convolutional Neural Networks: Influence and Implications of Target Model's Complexity}
  pdfauthor={Mathias M. P. Parisot, Balazs Pejo, Dayana Spagnuelo}
}

\newcommand{\eg}{\emph{e.g.},\xspace}
\newcommand{\ie}{\emph{i.e.},\xspace}

\author{Mathias P. M. Parisot\thanks{University of Amsterdam, Ansterdam, The Netherlands}, 
	Bal\'azs Pej\'o\thanks{CrySyS Lab, Dept. of Networked Systems and Services, Fac. of Electrical Engineering and Informatics, Budapest Univ. of Technology and Economics, Budapest, Hungary} and 
	Dayana Spagnuelo\thanks{Vrije Universiteit, Amsterdam, The Netherlands}}

\title{Property Inference Attacks on Convolutional Neural Networks:\\ Influence and Implications of Target Model's Complexity}
\date{}

\begin{document}

\maketitle

\subsubsection*{Abstract.}
Machine learning models' goal is to make correct predictions for specific tasks by learning important properties and patterns from data. By doing so, there is a chance that the model learns properties that are unrelated to its primary task. Property Inference Attacks exploit this and aim to infer from a given model (\ie the target model) properties about the training dataset seemingly unrelated to the model's primary goal. If the training data is sensitive, such an attack could lead to privacy leakage.
In this paper, we investigate the influence of the target model's complexity on the accuracy of this type of attack, focusing on convolutional neural network classifiers. We perform attacks on models that are trained on facial images to predict whether someone's mouth is open. Our attacks' goal is to infer whether the training dataset is balanced gender-wise. Our findings reveal that the risk of a privacy breach is present independently of the target model's complexity: for all studied architectures, the attack's accuracy is clearly over the baseline. We discuss the implication of the property inference on personal data in the light of Data Protection Regulations and Guidelines. 

\section{Introduction}

Machine Learning (ML) applications have received much attention over the last decade, mostly due to their vast application range, such as recommendation services, medicine, speech recognition, banking, gaming, driving, and more.
It is generally accepted that data plays a vital role in ML models' performance, and that more elaborate models can solve difficult tasks more accurately as they can learn more complex patterns from data. Notwithstanding, besides improving the performance, such ML models introduce privacy issues of the underlying datasets \cite{He2019}.




Training ML models typically requires significant amounts of data, potentially private and sensitive data, and the risk of privacy leakage is not negligible. Suppose a classification model that, once trained, can recognize the appropriate class of a data instance by learning mapping patterns between the training dataset and its original set of classes. This mapping is contained within the model parameters, for instance, in a neural network the mapping is encoded in each neuron's weights. 
As ML models are not exempt from malicious activity, an attacker knowing the trained model parameters could also gain some information about the data it was trained on. This is the rationale of Property Inference Attacks (PIAs) \cite{Ganju2018,Melis2019}, which aim at uncovering properties of the dataset in which a given model was trained by analyzing the parameters of the model only.

In a world where data has become a commodity---with increased amounts of data generated about people and increased willingness from companies to use such data to gain insights and improve their processes---, regulations come in place to ensure people's fundamental rights to privacy and control over personal data. An example of such is the European General Data Protection Regulation (GDPR)\footnote{Regulation (EU) 2016/679 of the European Parliament and of the Council of 27 April 2016 on the protection of natural persons with regard to the processing of personal data and on the free movement of such data, and repealing Directive 95/46/EC.}. 
The GDPR demands that personal data be processed ``lawfully, fairly and in a transparent manner'', and that principles of data minimization, integrity, and confidentiality be applied\footnote{GDPR, Art. 5}.

Given the current popularity and the increase in performance of ML applications, it is reasonable to question to which extent the training dataset is vulnerable to privacy attacks and to which extent the processing this data undergoes (e.g., used for ML training) is in line with the current data protection regulations and best practices. 
In particular, if improving a model's performance is realized by increasing the complexity of the model. As more complex models have more parameters and can retain more information about the training dataset, 
intuitively one may think that due to this information retention, more complex models could be more sensitive to PIAs as well. In this paper, we study this phenomenon, which---if the model is trained on a dataset containing personal data---could lead to potential privacy leakage.

\subsection{Contribution}

In this work, we test the influence of a model's complexity on its vulnerability to PIAs. Our setting is processing facial images; thus, we focus on Convolutional Neural Networks (CNNs), the most common model of choice for computer vision tasks. Given that we measure complexity in terms of the number of layers and weights of the model's architecture, \emph{we hypothesize that more complex models are also more sensitive to PIAs.} To uncover whether and how the risk of privacy leaks is influenced by the architecture's complexity of the model, we experiment on nine different CNN architectures. For each, we conduct 30 attacks based on 1500 shadow models.

\subsection{Organization}

In what follows, we first present works related to ours (Section \ref{sec:rw}), then we describe the methodology we follow to compose our attack (Section \ref{sec:metho}). In Section \ref{sec:experiment} we describe the implementation of our experimental setup, while in Section \ref{sec:disc} we present the results of our attack and discuss their meaning. Finally, in Section \ref{sec:conc}, we present final remarks regarding our findings' legal and ethical implications to data protection, and present future directions for our work.
\section{Related Works}
\label{sec:rw}


Several security threats are studied regarding machine learning focusing on the basic information security triad: confidentiality, integrity, and availability. For instance \cite{shumailov2020sponge} present an ML attack targeting the model's availability. According to \cite{He2019} the main attack categories for integrity are adversarial and poisoning attacks, while for confidentiality, these are model extraction and model inversion.

Adversarial attacks \cite{goodfellow2014explaining, szegedy2013intriguing} aim to take advantage of the weaknesses of the target model's classification boundary to craft data instances that are wrongly classified. 
Poisoning attacks \cite{mei2015using, jagielski2018manipulating} is similar to adversarial attacks as their goal is to influence the prediction of the target model. They do that by polluting the training set with malicious samples. While those are realistic threats to ML models' integrity, they do not pose immediate risk to data privacy. Instead, our work focuses on confidentiality attacks.

Concerning confidentiality (and privacy), the recent survey \cite{rigaki2020survey} gives a comprehensive overview of the subject. Here we only briefly comment on the most relevant works for our purposes,  we refer the reader to that work for more details. Model extraction \cite{tramer2016stealing, papernot2017practical, wang2018stealing} attacks aim at inferring the behavior of the target model to create a substitute model. Model inversion attacks \cite{fredrikson2015model,mehnaz2020black} aim at inferring information about the training data, for example, by reconstructing a representative of a particular class of the training set. This type of attack works better when all the instances of a single class represent the same entity. For example, \cite{fredrikson2015model} train a model to recognize individuals' faces, where the data instances of a given class all represent the same individual. The authors show that it is possible to reconstruct an image of this person. In such a case  \cite{mehnaz2020black} showed that some subgroups are more vulnerable to this type of attack than others.

Depending on the goal of the attacker, we can classify three more attacks under the category of model inversion: 
membership inference, reconstruction attack, and property inference attack. Membership inference attacks \cite{truex2018towards,hitaj2017deep,murakonda2019ultimate} aim at determining whether a particular data instance was used for training. This is severe privacy issues when the instance directly maps to an identifiable individual, for instance, a medical records dataset. In contrast, as its name suggests, reconstruction attacks \cite{zhu2020deep,li2019quantification} take this one step further and are capable of recovering both the same training inputs and the corresponding labels. 

The last one is PIA, the subject of this study. It aims to infer some hidden properties of the dataset that are independent of any class characteristics, and are therefore not necessarily related to the main classification task. Such properties can be general statistics about the dataset or can reflect biases in the training data. 

\subsection{Property Inference Attacks}

There have been few studies on PIAs with limited results. According to \cite{He2019}, only four papers were published on model inversion attacks, and researchers have not yet entirely determined the vulnerability of neural network architectures to privacy attacks such as PIA. Recent works perform PIAs in a federated learning setting, which allows multiple participants (also called clients by some works) to train a standard model without the need to share data. After each round of training, only the weights and the gradients are exchanged, while data remains protected on the participants' premises, helping to tackle privacy issues. 

The work presented in \cite{Melis2019} manages to infer properties that hold for a subset of the training data and that are independent of the property the target model aims to predict. Since the attack is performed during the training phase, it requires the model updates that are exchanged between participants. In contrast, the attack we focus on does not require the gradient values after each training round. We also target properties that are true for the whole dataset and not only for a subset. 

In \cite{Wang2019} three kinds of PIAs are proposed: class sniffing, quantity inference, and whole determination. Class sniffing detects whether a training label is present within a training round. Quantity inference determines how many clients have a given training label in their dataset. The whole determination infers the global proportion of a specific label. All of those attacks extract properties related to classification labels, and therefore to the main classification task. We focus on properties that are, in theory, unrelated to the task of the target model.

Attempts to explore user-level privacy leakage in a federated learning scenario is also subject of recent works \cite{wang2019beyond,pejo2020good}. They define client-dependent properties to precisely characterize the clients and distinguish them from each other. Both works present an attack that can uncover these differences. In more details, \cite{wang2019beyond} assume a malicious server with access to the individual updates and utilized Generative Adversarial Networks. In contrast, \cite{pejo2020good} assume an honest-but-curious setting and recover the participants' quality information via a differential attack without extra computational needs or access to the individual gradient updates. 

\subsection{Attacks Concerning Model Complexity}

A model inversion attack is presented in \cite{Zhang2019}. The authors study and theoretically prove the attack's relation with the model predictive power: more complex models, which should have greater predictive power, should also be more sensitive to model inversion attacks. However, the result of \cite{Zhang2019} was not proven for PIAs. We also focus more specifically on the complexity of the target model.

Model inversion attacks are also studied in recent works \cite{Geiping2020}. This work analyzes the effects of the target model's architecture on the difficulty of reconstructing input images. The authors investigate attacks on networks with various widths and depths and found that the width has the most significant influence on the reconstruction's quality. Contrary to ours, their study does not consider PIAs and is restricted to federated learning as their attack utilizes the gradient values.

In \cite{Ateniese2015} the first PIA attack using meta-classifiers is described, this is the methodology of the attack we use in our paper. Their research does not focus on the privacy leakage caused by such an attack but instead on the impact of the training set properties on the model performance. Moreover, they attack models implemented via Support Vector Machines and Hidden Markov Models using a binary tree meta-classifier but do not experiment with deep neural network models. 

Finally, an extension of the previous work \cite{Ateniese2015} is presented by \cite{Ganju2018}. The authors focus on neural networks and notice that a limitation of PIA performance is due to a property of fully connected networks called invariance. They propose two successful strategies to reduce this: converting a neural network to a canonical form, and using a deep set architecture. They use a pre-trained network to generate an embedding, which they feed as input to their target neural network. They perform the attack using the weights of the fully connected network following the pre-trained one. However, they do not study the influence of the type of layers and the model's architecture on the attack performance, which is the goal of our work.
\section{Methodology}
\label{sec:metho}

This section presents the attack strategy considered for our PIA alongside the assumptions about the target model. To improve readability, we summarize the paper's notations in Table \ref{tab:param}.

\begin{table}[h]
    \centering
    \caption{Notations used in the rest of the paper.}
    \begin{tabular}{cp{5cm}}
        \toprule
        Notation & Meaning \\
        \midrule
        $M_t$ & Target model (CNN)\\
        $D_t$ & Training dataset of the target model \\
        $P$ & Property to be inferred \\
        $M_{s_1},\dots,M_{s_k}$ & Shadow models, mimicking $M_t$\\
        $W_{s_1},\dots,W_{s_k}$ & Weights of the shadow models \\
        $D_{s_1},\dots,D_{s_k}$ & Training dataset of the shadow models \\
        $D$ & The dataset from which $D_{s_i}$ is created \\
        $M_a$ & Attack model to predict $P$ about $D_t$ \\
        $D_a$ & Training dataset of the attack model composed by $W_{s_1},\dots,W_{s_k}$ \\
        \bottomrule
    \end{tabular}
    \label{tab:param}
\end{table}

\subsection{Threat model}

Our target model is a CNN classifier. We assume a training dataset for the classifier, which contains sensitive data according to the definition by the GDPR\footnote{GDPR, Art. 9}, \eg data revealing racial or ethnic origin, religious beliefs, or biometric data. We assume an attacker whose goal is to infer general information about the training dataset, such as the proportion of the training data having a property $P$. This property is unrelated to the main classification task of the model. We assume the attacker can fabricate datasets similar to the original training dataset, \eg he or she knows from which distribution the original data was created. Moreover, the attacker can manipulate these datasets so that they either contain or not property $P$. We also assume the attacker has access to the target model and can train a large number of neural networks.

We focus on the white-box setting, where the attacker has access to the target model's full architecture and parameter values. Such information could be obtained in many ways: for instance, it could be shared explicitly as in Federated Learning \cite{shokri2015privacy}. In this scenario, the participants' datasets are hidden from each other (and from the aggregator); however, the participants and the aggregator server have full knowledge of the model and its parameters. Alternatively, when this information is not shared for neural networks, it could still be obtained by creating a substitute model with a similar decision boundary as the target model using a model extraction attack \cite{DBLP:journals/corr/PapernotMGJCS16}. 

\subsection{Attack}

We focus on PIAs whose goal is to extract general information about the target model's training dataset. This information is represented by a property $P$, which can be true or false. For instance, if the used dataset contains images of faces, $P$ can be defined as \textit{more than 20\% of the images within the dataset depict non-white people}. 

In this sense, PIA is transformed into a classification problem (for a given model) to determine whether it was trained on a dataset with property $P$. As such, the attack model can be understood as a meta-classifier as the dataset on which it is trained composed of shadow models, which are themselves classifiers. 
The attack consists of training $k$ shadow models ($M_{s_1},\dots,M_{s_k}$) on $k$ datasets ($D_{s_1},\dots,D_{s_k}$) specifically crafted to contain or not the target property $P$.
The training set ($D_a$) for the attack model ($M_a$) is composed by the weights ($W_{S_1},\dots,W_{S_k}$) of shadow models ($M_{s_1},\dots,M_{s_k}$) fabricated with the same architecture as the target model $M_t$.  

\begin{figure}[ht]
  \centering
  \includegraphics[width=\linewidth]{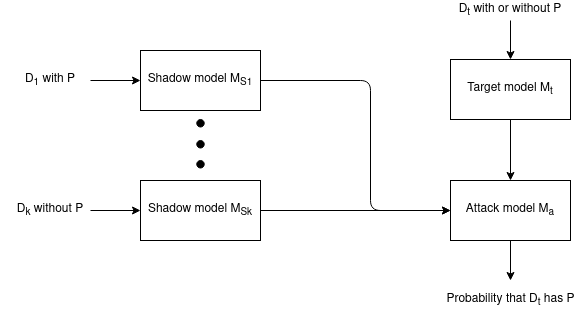}
  \caption{Property Inference Attack using a meta-classifier $M_a$ and datasets of shadow models $\{M_{s_1},\dots,M_{s_k}\}$.}
  \label{pia_diagram}
\end{figure}

The general overview of our attack is described in Figure \ref{pia_diagram}: we train an attack model that takes as input the weights of the target model and outputs the probability the dataset used for training the target model has property $P$ or not. This is based on the baseline PIA presented in \cite{Ganju2018}, which serves our purpose as we do not focus on the PIA itself but rather on the PIA's behavior performed on models with different complexities. 

\section{Experimental Setup}
\label{sec:experiment}


The experiments were performed on a laptop with an Intel i7-8750H (2.20GHz), 8GB RAM, and an Nvidia Quadro P600 GPU. The operating system is Ubuntu 20.04. The shadow models' training and the attack models were both done using Pytorch and are available on a public repository\footnote{The exact source will be shared after the blind review process to maintain anonymity.}.

\subsection{Datasets}

\begin{table*}[!t]
\centering
\caption{Layer-level description of each target \& shadow architectures. The detailed description of the parameters in each layer is presented in Table \ref{layer_description}.}
\resizebox{\columnwidth}{!}{%
\begin{tabular}{@{}lccccccccc@{}}
\toprule
    & Conv. 1               & Max-pool             & Conv. 2               & Max-pool             & Conv. 3               & Max-pool             & FC 1                 & FC 2                 & FC 3                 \\ \midrule
    & \multicolumn{1}{l}{} & \multicolumn{1}{l}{} & \multicolumn{1}{l}{} & \multicolumn{1}{l}{} & \multicolumn{1}{l}{} & \multicolumn{1}{l}{} & \multicolumn{1}{l}{} & \multicolumn{1}{l}{} & \multicolumn{1}{l}{} \\
$A_1$ & \checkmark           & \checkmark           & \checkmark           & \checkmark           & \checkmark           & \checkmark           & \checkmark           & \checkmark           & \checkmark           \\
$A_2$ & \checkmark           & \checkmark           & \checkmark           & \checkmark           & \checkmark           & \checkmark           & \checkmark           &                      & \checkmark           \\
$A_3$ & \checkmark           & \checkmark           & \checkmark           & \checkmark           & \checkmark           & \checkmark           &                      &                      & \checkmark           \\
$A_4$ & \checkmark           & \checkmark           & \checkmark           & \checkmark           &                      &                      & \checkmark           & \checkmark           & \checkmark           \\
$A_5$ & \checkmark           & \checkmark           & \checkmark           & \checkmark           &                      &                      & \checkmark           &                      & \checkmark           \\
$A_6$ & \checkmark           & \checkmark           & \checkmark           & \checkmark           &                      &                      &                      &                      & \checkmark           \\
$A_7$ & \checkmark           & \checkmark           &                      &                      &                      &                      & \checkmark           & \checkmark           & \checkmark           \\
$A_8$ & \checkmark           & \checkmark           &                      &                      &                      &                      & \checkmark           &                      & \checkmark           \\
$A_9$ & \checkmark           & \checkmark           &                      &                      &                      &                      &                      &                      & \checkmark           \\
    & \multicolumn{1}{l}{} & \multicolumn{1}{l}{} & \multicolumn{1}{l}{} & \multicolumn{1}{l}{} & \multicolumn{1}{l}{} & \multicolumn{1}{l}{} & \multicolumn{1}{l}{} & \multicolumn{1}{l}{} & \multicolumn{1}{l}{} \\ \bottomrule
\end{tabular}}
\label{shadow_architecture}
\end{table*}

To train the shadow models for our experiments, we have selected CelebFaces Attributes (CelebA) \cite{liu2015faceattributes} dataset, which contains personal and sensitive information. This is a face attributes dataset containing more than 200.000 face-centered images of 64 by 64 pixels of more than 10.000 celebrities. The images are labeled using 40 physical attributes such as hair color, smiling, and wearing a hat. Our shadow models and the target model are trained to detect whether the person appears with their mouth open in a given photo. Hence, we use the \textit{Mouth\_Open} attribute of the dataset.

Although this might seem like an irrelevant classification task, at the time of execution of this work, the world is undergoing a pandemic, and mouth covering masks is one of humanity's currently available weapons against the SARS-CoV-2 virus \cite{eikenberry2020mask}. This task can be related to automated mask detection\footnote{\url{http://tinyurl.com/2a8ewzvl}}, especially since, in many places, their use is compulsory. 

\subsection{Shadow Models}

The shadow models $\{M_{s_1},\dots,M_{s_k}\}$ are trained to mimic the target model $M_t$, \ie to differentiate between images of persons with and without their mouth open. However, the attacker's goal is to infer whether the training set of a given model was composed of an unbalanced number of images of males. This targeted property concerns biometric data and is classified as sensitive data according to the GDPR. In the real world, although it is not common, it is still possible to have gender-unbalanced generations in some societies: in the next 15 years in large parts of China and India, it is estimated that there will be around 20\% excess of young men \cite{hesketh2012effects}, which corresponds to 60\%-40\% division. Consequently, we set our thresholds to be at least double this gap: \textit{$P$ is true when the model's training set is composed of 70\% or more images containing males}. It is important to note that $P$ is not related to the model's classification task and that the target model does not use, at any time during training, the gender attribute.

The shadow models $\{M_{s_1},\dots,M_{s_k}\}$ have the same architecture as the targeted model and are trained to a reasonable level of accuracy to mimic the target model's behavior. We set the lowest acceptable accuracy to be 85\% (on the mouth open classification task) when the baseline distribution of the dataset is 51.7\%. For the attack to be effective, many shadow models have to be trained, as the attack model's inputs are the weights of the shadow models. Hence, for a specific target model architecture, we train 1800 shadow models. Since the computational cost of training many shadow models is significant, we decide not to use all images of \textit{CelebA}. Rather, for each shadow dataset $\{D_{s_1},\dots,D_{s_k}\}$, we use only 2000 randomly selected images. For half of the shadow models (i.e., 900 times) these 2000 images were selected to have property $P$, while the remaining 900 datasets do not. In detail, the exact proportion of males for each dataset was randomly taken from a uniform distribution either above or below 70\%, respectively. It is important to note that while each shadow model is trained using only 2000 images, they perform with at least 85\% accuracy on the whole test set of CelebA; therefore, they do not overfit to their small training set.

We experimented on target model's (and consequently the shadow models') architectures composed of up to 9 layers, each of three kinds: convolution layers, pooling layers, and fully connected layers. We trained 9 architectures ($A_1, \dots, A_9$) which are presented in Table \ref{shadow_architecture}, while the description of each layer is presented in Table \ref{layer_description}. The models take as input 64 by 64 RGB face images and output each picture's probability of representing a person with mouth open. Every architecture comprises 1-3 convolution layers, each followed by a max-pooling layer with a ReLU activation and 1-3 fully connected layers with a ReLU activation. The shadow models are trained for 50 epochs using the Mean Squared Error loss and the Adam optimizer with a learning rate of 0.001 and without decay or regularization. 

\begin{table}[h]
    \centering
    \caption{Description of the different layers used in the target \& shadow architectures.}
    \begin{tabular}{@{}ll@{}}
    \toprule
    Layer              & Description       \\ \midrule
    Convolution 1     & 6 filters 5x5     \\
    Max-pool          & 2x2, ReLU         \\
    Convolution 2     & 16 filters 5x5    \\
    Max-pool          & 2x2, ReLU         \\
    Convolution 3     & 32 filters 5x5    \\
    Max-pool          & 2x2, ReLU         \\
    Fully-Connected 1 & 120 neurons, ReLU \\
    Fully-Connected 2 & 84 neurons, ReLU  \\
    Fully-Connected 3 & 1 neuron          \\ \bottomrule
    \end{tabular}
    \label{layer_description}
\end{table}

\subsection{Attack Model and Evaluation}

The attack model classifies shadow models on whether they were trained on a dataset with or without the property $P$. The dataset used for the attack is composed of the 1800 shadow models and is split into training (1500 shadow models), validation (100 models), and test sets (200 models). The training algorithm is presented in Algorithm \ref{attack_algo}. The attack model is a simple multi-layer perceptron tuned using the validation set and evaluated on the test set.

\begin{algorithm}
\caption{Attack model training}\label{attack_algo}
\begin{algorithmic}[1]
    \Procedure{Train\_Attack}{$D, k$}
    \State{$D$ dataset of images, $k$ number of shadow models to train}
    \State{\textbf{let:} $D_{s_i}$ be a subset of $D$}
    \State{\textbf{let:} $D_{a}$ be the dataset used to train the attack model}
    \State{\textbf{let:} $P_{s_i}$ be a boolean value determining whether P is True on $D_{s_i}$}
    \State $D_a\gets \{\}$
    \For{$i \gets 1,k$}
    \State{$D_{s_i} \gets$ sample subset of $D$}
    \State{$P_{s_i} \gets$ evaluate P on $D_{s_i}$}
    \State{$M_{s_i} \gets train(D_{s_i})$}
    \State{$W_{s_i} \gets getWeights(M_{s_i})$}
    \State{$D_a \gets D_a \cup \{(W_{s_i}, P_{s_i})$\}}
    \EndFor
    \State{$M_a \gets train(D_a)$}
    \State \textbf{return} $M_a$
    \EndProcedure
\end{algorithmic}
\end{algorithm}

We tuned the attack model by performing a grid search over the following hyper-parameters: learning rate, loss function, batch size, optimizer, the activation function of the first layer of the attack model. The hyper-parameters values we used are presented in Table \ref{hyperparameters}. We trained six attack models for the 9 model architectures for each of the combinations of parameters during ten epochs. We selected the best parameters by combining the largest median accuracy over the 9 model architectures. 

\begin{table}[h]
    \centering
    \caption{Attack model hyper-parameter tuning with the optimal ones marked in bold.}
    \begin{tabular}{ll}
    \toprule
    Parameter              & Values       \\ \midrule
    Learning rate & \textbf{0.005}; 0.001; 0.0005  \\
    Loss function & \textbf{MSE}; L1-loss \\ 
    Batch size & 16; \textbf{32}; 64 \\
    Optimizer & SGD; \textbf{Adam} \\ 
    Input layer act. func. & sigmoid; \textbf{ReLU}; tanh \\ \bottomrule
    \end{tabular}
    \label{hyperparameters}
\end{table}

After tuning, the selected attack model architecture is presented in Table \ref{attack_architecture}. The attack model's inputs are the flattened weights of the model it is trying to classify as having or not the property $P$. Therefore, a target model architecture with a larger number of parameters induces a more comprehensive input layer for the attack model. The attack model comprises two fully connected layers, which we trained 30 times for each shadow model architecture. We presented the average performances after 20 epoch when the training was done using the Mean Squared Error loss function and the Adam optimizer with a learning rate of 0.005 and without decay or regularization. 

\begin{table}[h]
    \centering
    \caption{Architectures of the attack model. FC1 is the input layer and FC2 the output layer.}
    \begin{tabular}{@{}ll@{}}
    \toprule
    Name              & Description       \\ \midrule
    Fully-Connected 1 & 10 neurons, ReLU  \\
    Fully-Connected 2 & 1 neuron          \\ \bottomrule
    \end{tabular}
    \label{attack_architecture}
\end{table}
\section{Results and Discussions}
\label{sec:disc}

\begin{figure*}[ht]
     \centering
     \begin{subfigure}{0.49\textwidth}
         \centering
         \includegraphics[width=\textwidth]{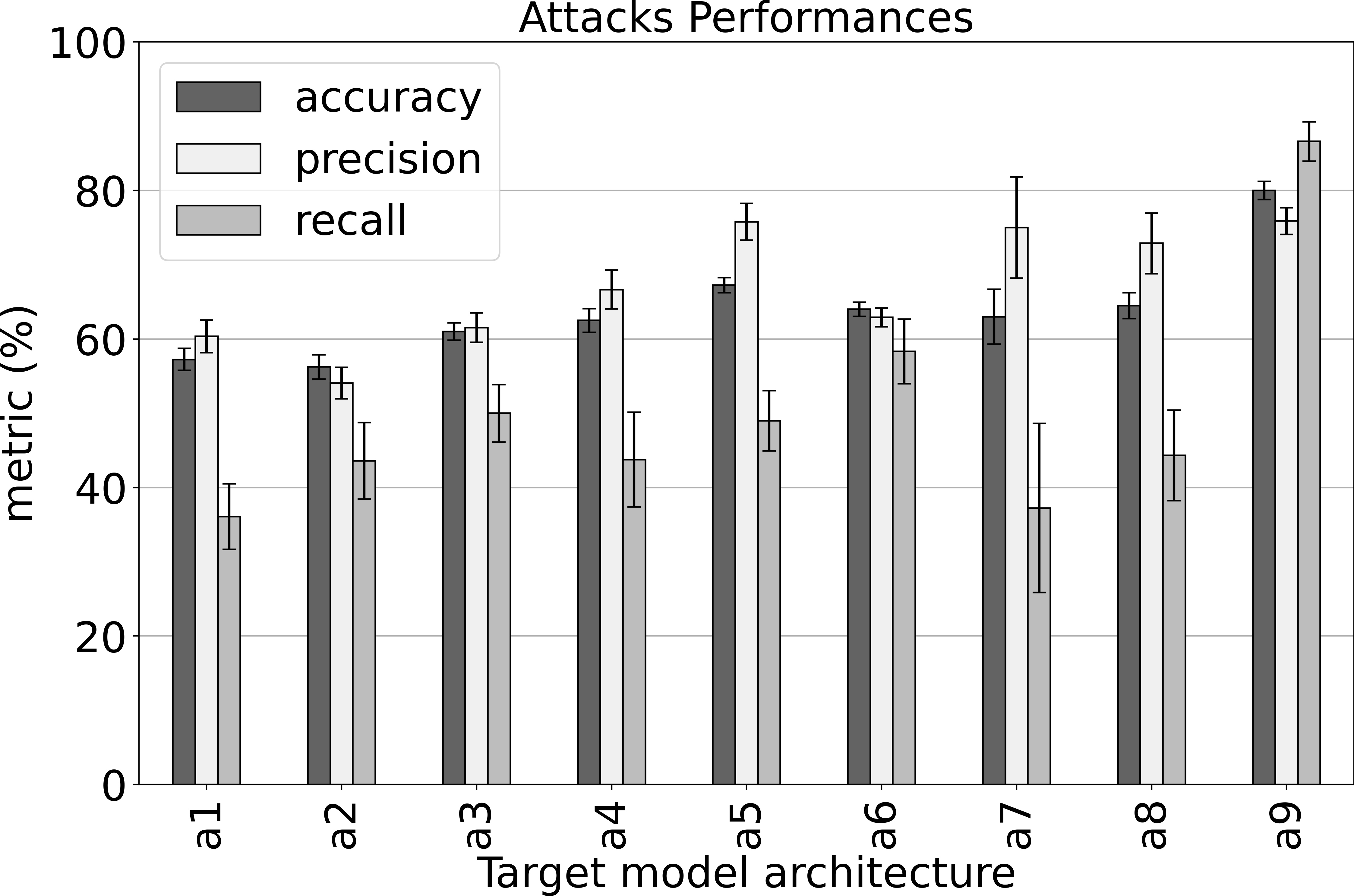}
         \caption{Attack's accuracy, precision, and recall using weights from entire target model.}
         \label{accuracy_perf}
     \end{subfigure}
     \hfill
     \begin{subfigure}{0.49\textwidth}
         \centering
         \includegraphics[width=\textwidth]{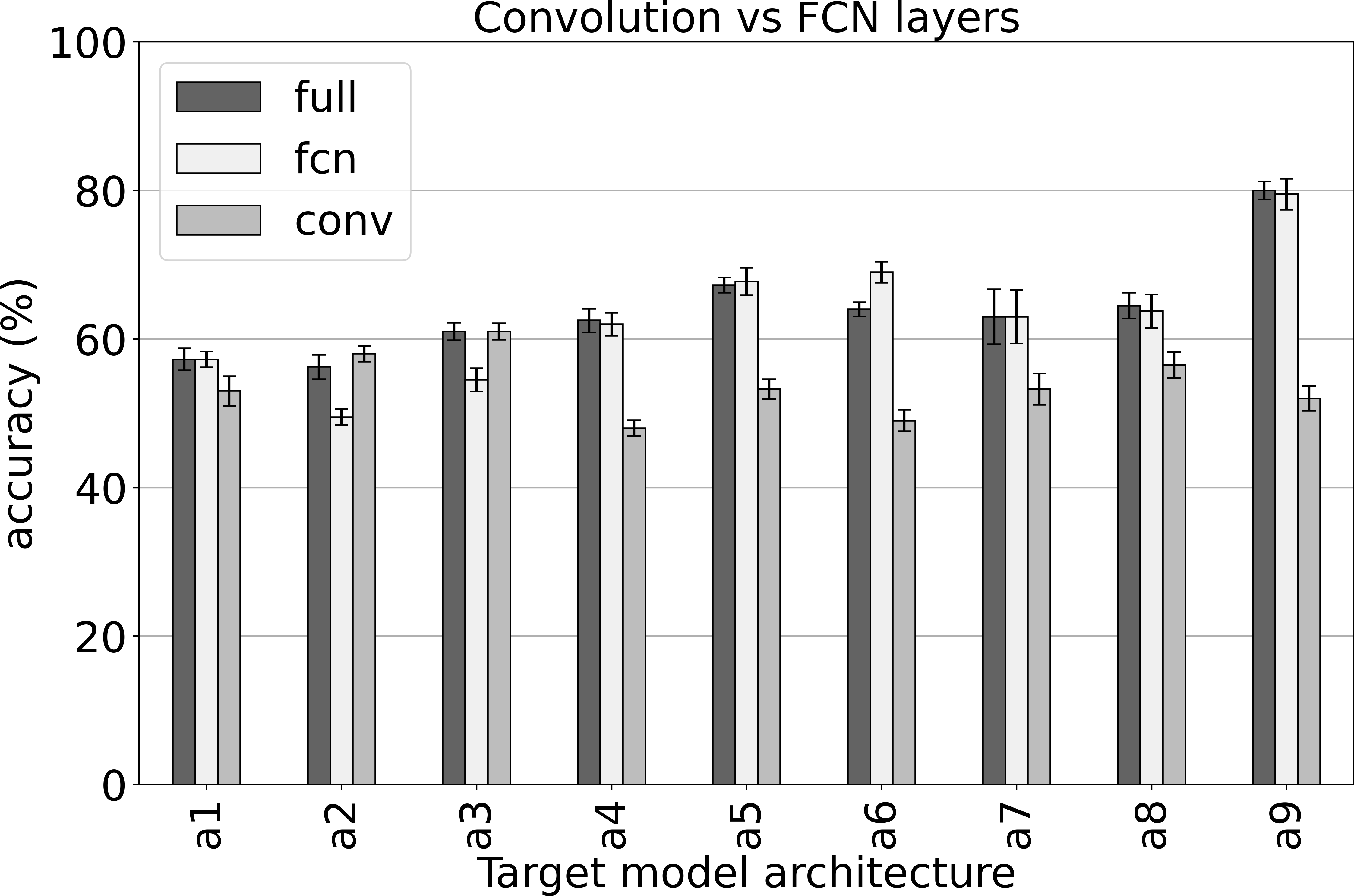}
         \caption{Attack's accuracy using weights from: entire target model (full), only fully connected layers (fcn), and only convolution layers (conv).}
         \label{accuracy_conv_fcn}
     \end{subfigure}
    \caption{Attack's performance on each architecture. Bars correspond to the median of 30 attacks, and error bars to $\pm$ the standard deviation.}
    \label{accuracy}
\end{figure*}

Figure \ref{accuracy} summarizes the performance of our attack. In detail, Figure \ref{accuracy_perf} presents the accuracy of the attacks on each target model architecture, which varies between $56\%-80\%$ depending on the architecture. These results confirm the findings of \cite{Ganju2018} that the target models do learn information unrelated to the task they were trained to learn.

We create as many shadow models presenting the property $P$ as ones not presenting it in our setup. Therefore, the expected baseline is 50\% accuracy. Most of the architectures have less than 67\% attack accuracy, so the real privacy threat is low. 
However, the attack model was tuned to achieve good performance across the nine shadow model architectures. Fine-tuning the attack model for a specific architecture should improve the attack accuracy. Moreover, \cite{Ganju2018} have shown that it is possible to significantly increase the accuracy of this attack using representations that are invariant to node permutations. They managed to get almost perfect accuracy making the attack usable in a real-world setting. 

We performed PIAs on distinctive neural network architectures with different amounts and types of layers. As convolution layers and fully connected ones play different roles in a CNN, we also tested whether the type of used layers impacts the attack's accuracy. Thus we conducted three additional PIAs on each of the architectures presented in Table \ref{shadow_architecture}: 1) using all the weights of the shadow model; 2) using only the weights of the convolution layers, and 3) using only the weights of the fully connected layers. Figure \ref{accuracy_conv_fcn} presents the accuracy of the three attacks. For most target model architectures, the PIA using only the fully connected weights performs as well, and sometimes better, as the PIA using the weights from both types of layers. Consequently, the information leaked by a CNN seems to be contained in the fully connected part of the network.

\begin{figure}[ht]
    \centering
    \includegraphics[width=\linewidth]{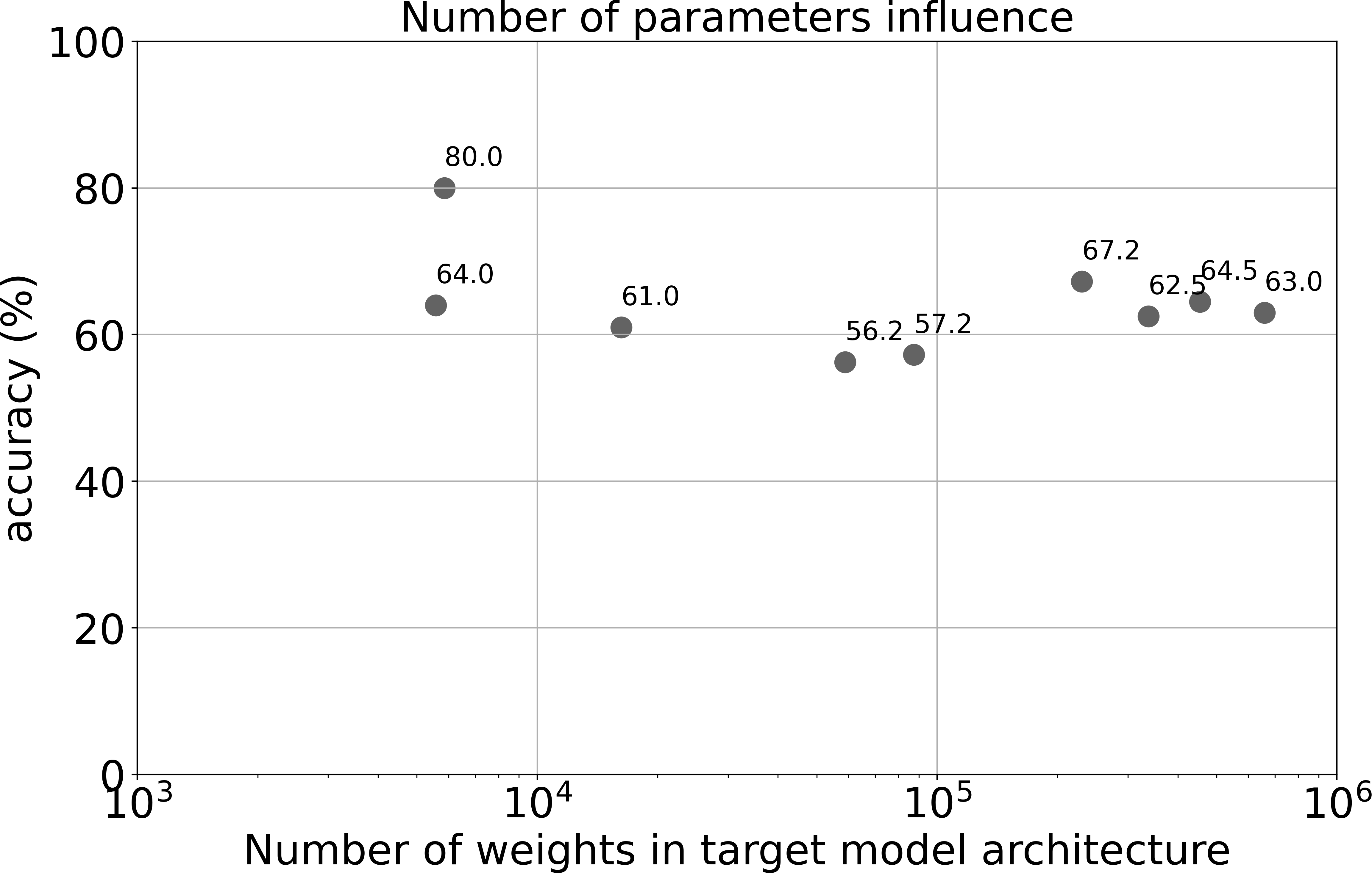}
    \caption{Influence of the complexity of the target model (express as the number of parameters) on the accuracy of the attacks on each architecture. Each dot corresponds to the median of 30 attacks.}
    \label{accuracy_weight}
\end{figure}

One of this study's goals is to establish whether there is a relationship between model complexity (which we define by the number of parameters) and its sensibility to PIAs defined as the accuracy of the attack. Figure \ref{accuracy_weight} summarizes this relationship. Our results do not directly support this claim.

\section{Conclusion}
\label{sec:conc}

This work presents an attack that tries to determine if a given dataset (of faces) used to train a CNN model (to determine if a mouth is open in a picture) has some property $P$, in our case, whether the dataset was unbalanced gender-wise. We conducted several experiments to uncover the relationship between target model complexity and privacy leakage (aka PIA's accuracy), and we find no significant correlation between the two. 

Due to the nature of PIAs, our work has an explicit limitation: it is tailored to one specific property of the dataset. Our attack can be adapted to other properties $P$ as well, as long as the attacker can fabricate datasets containing or not the given property. 
It is possible, for instance, to infer whether a training set contains pictures of a given person. In this case, when the instances of the same class represent the same entity (same person), a PIA is transformed into a membership inference attack which is not only focused on a specific data instance but on whether the whole dataset contains a particular class of image (a particular person). 
This type of membership inference attack presents a distinct type of threat.
Regular membership inference attacks rely on the attacker to possess the same data instance present in the dataset. For instance, if the dataset is composed of portrait images, the attacker must possess the same image of a given person that belongs to the dataset.
It may not be realistic to assume that attackers will get hold of the very same image. However, it is more reasonable to expect an attacker can produce a dataset of different targeted persons' pictures. The attack then consists of creating shadow datasets with and without pictures of the targeted person and using them to train shadow models and perform a PIA.

\subsection{Data Protection Implications}

Our experimental attack highlights a surplus of personal and sensitive data for training the classification models (attacks' accuracy are equal or higher than 56\%). In the context of the current data protection regulation (\ie GDPR), this can be framed as a violation of the \textit{data minimization} principle of data processing, which explicitly requests that processing of personal data be ``adequate, relevant and limited to what is necessary concerning the purposes for which they are processed''\footnote{GDPR, Art. 5.1}. In this paper, the PIA demonstrates that for the task of classifying whether someone's mouth is open, there is a leak of the gender balance in the training dataset. This can be attributed to the fact that the dataset comprises full faces instead of only mouth images, which is arguably the only data indeed necessary for the task.

An intuitive solution for the surplus of data problem would be to crop images to contain only the area relevant for the classification task. In our example, a preliminary step identifies mouths in the images and eliminates the remaining of the face. However, this alone could prove insufficient in the context of the attack we present in this paper. As empirically demonstrated by Pew Research Center\footnote{https://www.pewresearch.org/interactives/how-does-a-computer-see-gender/}, gender traits can be encoded in many (unexpected) areas of a facial image. Recent works explore alternative techniques that could help achieve data minimization in images, such as feature anonymization \cite{kim2020selective}, and gender obfuscation through morphing \cite{wang2020gender}. A possible future research direction would be to rerun our experiment with models trained on datasets using those techniques and test their impact on PIAs.

The GDPR also demands that people be informed, among others, of the envisioned consequences of data processing, which leads to automated decision-making\footnote{GDPR, Art. 13.2(f)}. This request is part of the concept of transparency, which is essential to ensure that people can exercise their rights under the GDPR \cite{demetzou2018gdpr}.
Our PIAs brings to light one of the hidden threats to people's data due to such processing. However, it is reasonable to assume other types of attacks may uncover different threats.
While informing of possible risks seems fundamental to grant people's right to privacy, thoroughly doing so may require disproportional technical efforts from the data controller's side. For instance, requiring controllers to run experiments similar to the one presented in this paper. 
Some works in the literature seem to agree with the vision that risks and consequences of ``undesired outcomes'' from machine learning models are of interest to people whose data is processed \cite{ras2018explanation}. Nevertheless, the question that remains is to what extent should a data controller explore the possible risks to \textit{fairly} inform people of \textit{envisioned} consequences. Answering this question is a task we leave for future works.

More conservative school of thought argues that machine learning models--- specifically the ones for which privacy attacks are possible ---might be seen as a type of personal data themselves \cite{veale2018algorithms}. The authors argue that model inversion (of which PIA is a type, and aims at estimating the training data from a publicly available model), despite some level of uncertainty, leads to recovering of data that will be more accurate than simply guessing characteristics of the original training dataset. A comparison is made with `pseudonymization'\footnote{A note on terminology: the notion of `psuedonymization' as described by the GDPR resembles what works from the discipline of computer science refer to as `anonymization'.}, which aims at transforming data in a way that can no longer identify a person without the use of additional data\footnote{GDPR, Art. 4.5}. The authors present a legal precedent that shows this data need not be all in possession of the same entity for the re-identification to be considered a privacy breach (\textit{Breyer} (ECLI:EU:C:2016:779) as cited in \cite{veale2018algorithms}). In the same way, PIAs uncovering new data about the data set would also characterize a privacy breach, even if no personal data is made available by the data controllers themselves. 
This is turn triggers a series of rights for the data subjects (people whose data are in the dataset), and obligations for the data controller (organization training and using the model), among those the one of complying with the security and data protection by design principles. The conclusion is that sharing or making available models for which privacy attacks are possible, if done with no legal basis (which is not needed for anonymous or non-personal data), would be characterize a violation of such principles and a privacy breach.

Even if such arguments might feel unrealistic at the moment, we draw attention to guidelines recently published by public authorities suggesting the best practices for for Data Protection in Artificial Intelligence (AI).
They evidence that a requirement for actively thinking about the risks of PIAs and other types of attacks to machine learning models is in the horizon. The guidelines on AI and data protection by the Council of Europe call for \textit{algorithm vigilance} and suggest legislators and policy makers to require ``prior assessment of the impact of data processing on human rights and fundamental freedoms, and vigilance on the potential adverse effects and consequences of AI applications" (Section III.2 in \cite{AIandDP}). Other similar guidelines also emphasize the role of assessing the expected impacts and risks (Data Protection Impact Assessment) for the data subject in the beginning of an AI project \cite{declaration, AIandPrivacy}.


\paragraph{Concluding remarks.} In this work, we present an experimental setting to test the influence and implications of the target model's complexity in the accuracy of Property Inference Attacks (PIAs). Although our findings did not support our initial hypothesis that more complex models would intrinsically learn more information from the training dataset and hence be more sensitive to PIAs, they reveal a surplus of personal information used in the training stage of CNN models. The implications of our work is shown to have an impact on the rights and obligations with respect to Data Protection Regulations and Guidelines.

\bibliographystyle{apalike}
{\small
\bibliography{main}}

\begin{thebibliography}{}

\bibitem[Ateniese et~al., 2015]{Ateniese2015}
Ateniese, G., Mancini, L.~V., Spognardi, A., Villani, A., Vitali, D., and
  Felici, G. (2015).
\newblock Hacking smart machines with smarter ones: How to extract meaningful
  data from machine learning classifiers.
\newblock {\em International Journal of Security and Networks}, 10(3):137--150.

\bibitem[{Commission Nationale de l’Informatique et des Libertés (CNIL)}
  et~al., 2018]{declaration}
{Commission Nationale de l’Informatique et des Libertés (CNIL)}, {European
  Data Protection Supervisor (EDPS)}, and {Garante per la protezione dei dati
  personali} (2018).
\newblock Declaration on ethics and data protection in artificial intelligence.
\newblock
  \url{http://globalprivacyassembly.org/wp-content/uploads/2018/10/20180922_ICDPPC-40th_AI-Declaration_ADOPTED.pdf}.
\newblock Last accessed 25 February 2021.

\bibitem[{Council of Europe}, 2019]{AIandDP}
{Council of Europe} (2019).
\newblock Guidelines on artificial intelligence and data protection.
\newblock
  \url{https://rm.coe.int/guidelines-on-artificial-intelligence-and-data-protection/168091f9d8}.
\newblock Last accessed 25 February 2021.

\bibitem[Demetzou, 2018]{demetzou2018gdpr}
Demetzou, K. (2018).
\newblock {GDPR and the Concept of Risk}.
\newblock In {\em IFIP International Summer School on Privacy and Identity
  Management}, pages 137--154. Springer.

\bibitem[Eikenberry et~al., 2020]{eikenberry2020mask}
Eikenberry, S.~E., Mancuso, M., Iboi, E., Phan, T., Eikenberry, K., Kuang, Y.,
  Kostelich, E., and Gumel, A.~B. (2020).
\newblock To mask or not to mask: Modeling the potential for face mask use by
  the general public to curtail the covid-19 pandemic.
\newblock {\em Infectious Disease Modelling}.

\bibitem[Fredrikson et~al., 2015]{fredrikson2015model}
Fredrikson, M., Jha, S., and Ristenpart, T. (2015).
\newblock Model inversion attacks that exploit confidence information and basic
  countermeasures.
\newblock In {\em Proceedings of the 22nd ACM SIGSAC Conference on Computer and
  Communications Security}, pages 1322--1333.

\bibitem[Ganju et~al., 2018]{Ganju2018}
Ganju, K., Wang, Q., Yang, W., Gunter, C.~A., and Borisov, N. (2018).
\newblock Property inference attacks on fully connected neural networks using
  permutation invariant representations.
\newblock pages 619--633.

\bibitem[Geiping et~al., 2020]{Geiping2020}
Geiping, J., Bauermeister, H., Dr{\"o}ge, H., and Moeller, M. (2020).
\newblock Inverting gradients--how easy is it to break privacy in federated
  learning?
\newblock {\em arXiv preprint arXiv:2003.14053}.

\bibitem[Goodfellow et~al., 2014]{goodfellow2014explaining}
Goodfellow, I.~J., Shlens, J., and Szegedy, C. (2014).
\newblock Explaining and harnessing adversarial examples.
\newblock {\em arXiv preprint arXiv:1412.6572}.

\bibitem[He et~al., 2019]{He2019}
He, Y., Meng, G., Chen, K., Hu, X., and He, J. (2019).
\newblock Towards privacy and security of deep learning systems: a survey.
\newblock {\em arXiv preprint arXiv:1911.12562}.

\bibitem[Hesketh and Min, 2012]{hesketh2012effects}
Hesketh, T. and Min, J.~M. (2012).
\newblock The effects of artificial gender imbalance: Science \& society series
  on sex and science.
\newblock {\em EMBO reports}, 13(6):487--492.

\bibitem[Hitaj et~al., 2017]{hitaj2017deep}
Hitaj, B., Ateniese, G., and Perez-Cruz, F. (2017).
\newblock Deep models under the gan: information leakage from collaborative
  deep learning.
\newblock In {\em Proceedings of the 2017 ACM SIGSAC Conference on Computer and
  Communications Security}, pages 603--618.

\bibitem[Jagielski et~al., 2018]{jagielski2018manipulating}
Jagielski, M., Oprea, A., Biggio, B., Liu, C., Nita-Rotaru, C., and Li, B.
  (2018).
\newblock Manipulating machine learning: Poisoning attacks and countermeasures
  for regression learning.
\newblock In {\em 2018 IEEE Symposium on Security and Privacy (SP)}, pages
  19--35. IEEE.

\bibitem[Kim and Yang, 2020]{kim2020selective}
Kim, T. and Yang, J. (2020).
\newblock Selective feature anonymization for privacy-preserving image data
  publishing.
\newblock {\em Electronics}, 9(5):874.

\bibitem[Li et~al., 2019]{li2019quantification}
Li, Z., Huang, Z., Chen, C., and Hong, C. (2019).
\newblock Quantification of the leakage in federated learning.
\newblock {\em arXiv preprint arXiv:1910.05467}.

\bibitem[Liu et~al., 2015]{liu2015faceattributes}
Liu, Z., Luo, P., Wang, X., and Tang, X. (2015).
\newblock Deep learning face attributes in the wild.
\newblock In {\em Proceedings of International Conference on Computer Vision
  (ICCV)}.

\bibitem[Mehnaz et~al., 2020]{mehnaz2020black}
Mehnaz, S., Li, N., and Bertino, E. (2020).
\newblock Black-box model inversion attribute inference attacks on
  classification models.
\newblock {\em arXiv preprint arXiv:2012.03404}.

\bibitem[Mei and Zhu, 2015]{mei2015using}
Mei, S. and Zhu, X. (2015).
\newblock Using machine teaching to identify optimal training-set attacks on
  machine learners.
\newblock In {\em Twenty-Ninth AAAI Conference on Artificial Intelligence}.

\bibitem[Melis et~al., 2019]{Melis2019}
Melis, L., Song, C., De~Cristofaro, E., and Shmatikov, V. (2019).
\newblock Exploiting unintended feature leakage in collaborative learning.
\newblock pages 691--706.

\bibitem[Murakonda et~al., 2019]{murakonda2019ultimate}
Murakonda, S.~K., Shokri, R., and Theodorakopoulos, G. (2019).
\newblock Ultimate power of inference attacks: Privacy risks of learning
  high-dimensional graphical models.
\newblock {\em arXiv preprint arXiv:1905.12774}.

\bibitem[Papernot et~al., 2016]{DBLP:journals/corr/PapernotMGJCS16}
Papernot, N., McDaniel, P., Goodfellow, I., Jha, S., Celik, Z.~B., and Swami,
  A. (2016).
\newblock Practical black-box attacks against deep learning systems using
  adversarial examples.
\newblock {\em arXiv preprint arXiv:1602.02697}, 1(2):3.

\bibitem[Papernot et~al., 2017]{papernot2017practical}
Papernot, N., McDaniel, P., Goodfellow, I., Jha, S., Celik, Z.~B., and Swami,
  A. (2017).
\newblock Practical black-box attacks against machine learning.
\newblock In {\em Proceedings of the 2017 ACM on Asia conference on computer
  and communications security}, pages 506--519.

\bibitem[Pej{\'o}, 2020]{pejo2020good}
Pej{\'o}, B. (2020).
\newblock The good, the bad, and the ugly: Quality inference in federated
  learning.
\newblock {\em arXiv preprint arXiv:2007.06236}.

\bibitem[Ras et~al., 2018]{ras2018explanation}
Ras, G., van Gerven, M., and Haselager, P. (2018).
\newblock Explanation methods in deep learning: Users, values, concerns and
  challenges.
\newblock In {\em Explainable and Interpretable Models in Computer Vision and
  Machine Learning}, pages 19--36. Springer.

\bibitem[Rigaki and Garcia, 2020]{rigaki2020survey}
Rigaki, M. and Garcia, S. (2020).
\newblock A survey of privacy attacks in machine learning.
\newblock {\em arXiv preprint arXiv:2007.07646}.

\bibitem[Shokri and Shmatikov, 2015]{shokri2015privacy}
Shokri, R. and Shmatikov, V. (2015).
\newblock Privacy-preserving deep learning.
\newblock In {\em Proceedings of the 22nd ACM SIGSAC conference on computer and
  communications security}, pages 1310--1321.

\bibitem[Shumailov et~al., 2020]{shumailov2020sponge}
Shumailov, I., Zhao, Y., Bates, D., Papernot, N., Mullins, R., and Anderson, R.
  (2020).
\newblock Sponge examples: Energy-latency attacks on neural networks.
\newblock {\em arXiv preprint arXiv:2006.03463}.

\bibitem[Szegedy et~al., 2013]{szegedy2013intriguing}
Szegedy, C., Zaremba, W., Sutskever, I., Bruna, J., Erhan, D., Goodfellow, I.,
  and Fergus, R. (2013).
\newblock Intriguing properties of neural networks.
\newblock {\em arXiv preprint arXiv:1312.6199}.

\bibitem[{The Norwegian Data Protection Authority (Datatilsynet)},
  2018]{AIandPrivacy}
{The Norwegian Data Protection Authority (Datatilsynet)} (2018).
\newblock Artificial intelligence and privacy.
\newblock
  \url{https://www.datatilsynet.no/globalassets/global/english/ai-and-privacy.pdf}.
\newblock Last accessed 25 February 2021.

\bibitem[Tram{\`e}r et~al., 2016]{tramer2016stealing}
Tram{\`e}r, F., Zhang, F., Juels, A., Reiter, M.~K., and Ristenpart, T. (2016).
\newblock Stealing machine learning models via prediction apis.
\newblock In {\em 25th $\{$USENIX$\}$ Security Symposium ($\{$USENIX$\}$
  Security 16)}, pages 601--618.

\bibitem[Truex et~al., 2018]{truex2018towards}
Truex, S., Liu, L., Gursoy, M.~E., Yu, L., and Wei, W. (2018).
\newblock Towards demystifying membership inference attacks.
\newblock {\em arXiv preprint arXiv:1807.09173}.

\bibitem[Veale et~al., 2018]{veale2018algorithms}
Veale, M., Binns, R., and Edwards, L. (2018).
\newblock Algorithms that remember: model inversion attacks and data protection
  law.
\newblock {\em Philosophical Transactions of the Royal Society A: Mathematical,
  Physical and Engineering Sciences}, 376(2133):20180083.

\bibitem[Wang and Gong, 2018]{wang2018stealing}
Wang, B. and Gong, N.~Z. (2018).
\newblock Stealing hyperparameters in machine learning.
\newblock In {\em 2018 IEEE Symposium on Security and Privacy (SP)}, pages
  36--52. IEEE.

\bibitem[Wang et~al., 2019a]{Wang2019}
Wang, L., Xu, S., Wang, X., and Zhu, Q. (2019a).
\newblock Eavesdrop the composition proportion of training labels in federated
  learning.
\newblock {\em arXiv preprint arXiv:1910.06044}.

\bibitem[Wang, 2020]{wang2020gender}
Wang, S. (2020).
\newblock Gender obfuscation through face morphing.
\newblock Master's thesis, University of Twente.

\bibitem[Wang et~al., 2019b]{wang2019beyond}
Wang, Z., Song, M., Zhang, Z., Song, Y., Wang, Q., and Qi, H. (2019b).
\newblock Beyond inferring class representatives: User-level privacy leakage
  from federated learning.
\newblock In {\em IEEE INFOCOM 2019-IEEE Conference on Computer
  Communications}, pages 2512--2520. IEEE.

\bibitem[Zhang et~al., 2020]{Zhang2019}
Zhang, Y., Jia, R., Pei, H., Wang, W., Li, B., and Song, D. (2020).
\newblock The secret revealer: Generative model-inversion attacks against deep
  neural networks.
\newblock In {\em Proceedings of the IEEE/CVF Conference on Computer Vision and
  Pattern Recognition}, pages 253--261.

\bibitem[Zhu and Han, 2020]{zhu2020deep}
Zhu, L. and Han, S. (2020).
\newblock Deep leakage from gradients.
\newblock In {\em Federated Learning}, pages 17--31. Springer.

\end{thebibliography}

\end{document}